\documentclass[fleqn,usenatbib]{mnras}
\usepackage{newtxtext,newtxmath}
\usepackage[T1]{fontenc}

\DeclareRobustCommand{\VAN}[3]{#2}
\let\VANthebibliography\thebibliography
\def\thebibliography{\DeclareRobustCommand{\VAN}[3]{##3}\VANthebibliography}

\usepackage{graphicx}	% Including figure files
\usepackage{amsmath}	% Advanced maths commands
\usepackage{ulem}

\newcommand{\stageAsh}{0.059387(5)~} 
\newcommand{\stageBsh}{0.058864(3)~} 
\newcommand{\pdot}{$-$1.4(2)$\times10^{-5}$~}
\newcommand{\massratio}{0.054(7)~}
\newcommand{\risets}{1.62(9) d mag$^{-1}$~} %subset(ato, V1 > 60859 & V1 < 60869.5)
\newcommand{\plateuts}{35.2(1) d mag$^{-1}$~} %subset(dat, V1 > 60876.5 & V1 < 60899.6735)

\def\commenta{$^*$}
\def\commentb{$^\dagger$}
\def\commentc{$^\ddagger$}
\def\commentd{$^\S$}
\def\commente{$^\|$}

\usepackage{array}
\usepackage{xcolor}

\title[An inside-out outburst from a period bouncer]{A probable inside-out dwarf nova outburst from the period bouncer candidate ASASSN-25dc}

\author[Y. Tampo et al.]{
Yusuke Tampo,$^{1,2}$\thanks{E-mail: yusuke@saao.ac.za (YT)}
Naoto Kojiguchi,$^{3,4}$
Mariko Kimura,$^{5}$
Keisuke Isogai,$^{3,6}$ 
David. A. H. Buckley,$^{1,2,7}$
\newauthor
Nikita Rawat,$^{1,}$
Stephen B. Potter,$^{1,8}$
Anke van Dyk,$^{1,2}$
Patrick Woudt,$^{2}$ 
Paul J. Groot,$^{1,2,9}$
\newauthor
Franz-Josef Hambsch,$^{10,11,12}$ 
Berto Monard,$^{13}$ 
Peter Starr,$^{14}$ 
William Goltz,$^{15}$ 
Daisaku Nogami,$^{4}$ 
\newauthor
and 
Taichi Kato$^{4}$ 
\\
$^{1}$South African Astronomical Observatory, PO Box 9, Observatory, 7935, Cape Town, South Africa\\
$^{2}$Department of Astronomy, University of Cape Town, Private Bag X3, Rondebosch 7701, South Africa\\
$^{3}$Okayama Observatory, Kyoto University, 3037-5 Honjo, Kamogatacho, Asakuchi, Okayama 719-0232, Japan\\
$^{4}$Department of Astronomy, Kyoto University, Kitashirakawa-Oiwake-cho, Sakyo-ku, Kyoto 606-8502, Japan\\
$^{5}$Advanced Research Center for Space Science and Technology, Colledge of Science and Engineering, Kanazawa University, \\Kakuma, Kanazawa, Ishikawa 920-1192, Japan\\
$^{6}$Department of Multi-Disciplinary Sciences, Graduate School of Arts and Sciences, The University of Tokyo, 3-8-1 Komaba, Meguro, Tokyo 153-8902, Japan\\
$^{7}$Department of Physics, University of the Free State, P.O. Box 339, Bloemfontein 9300, South Africa\\
$^{8}$Department of Physics, University of Johannesburg, PO Box 524, Auckland Park 2006, South Africa\\
$^{9}$Department of Astrophysics/IMAPP, Radboud University, PO Box 9010, 6500 GL Nijmegen, The Netherlands\\
$^{10}$Groupe Européen d’Observations Stellaires (GEOS), 23 Parc de Levesville, 28300 Bailleau l’Evêque, France\\
$^{11}$Bundesdeutsche Arbeitsgemeinschaft für Veränderliche Sterne (BAV), Munsterdamm 90, 12169 Berlin, Germany\\
$^{12}$Vereniging Voor Sterrenkunde (VVS), Zeeweg 96, 8200 Brugge, Belgium\\
$^{13}$Kleinkaroo Observatory, Center for Backyard Astrophysics Kleinkaroo, Sint Helena 1B, PO Box 281, Calitzdorp 6660, South Africa\\
$^{14}$Dubbo Observatory, 17L Camp Rd, Dubbo NSW 2830, Australia\\
$^{15}$Association of Variable Star Observers (AAVSO), 185 Alewife Brook Parkway, Suite 410, Cambridge, MA 02138, United States of America\\
}

\date{Accepted XXX. Received YYY; in original form ZZZ}

\pubyear{\the\year{}}

\begin{document}
\label{firstpage}
\pagerange{\pageref{firstpage}--\pageref{lastpage}}
\maketitle

\begin{abstract}

We report optical time-resolved photometric observations of a newly-discovered outbursting system, ASASSN-25dc. Its 8-mag amplitude, 40-day duration, 1-mag dip in the outburst plateau, and positive superhumps are characteristic of a dwarf nova superoutburst in a non-magnetic cataclysmic variable. We establish its stage-A and stage-B superhump periods as \stageAsh d and \stageBsh d, respectively. The negative superhump period derivative (\pdot cycle$^{-1}$) during the stage-B superhumps and the empirical relation indicate the mass ratio is \massratio, below the period bounce range. The long outburst decline timescale (\plateuts) and small superhump amplitude ($\simeq$0.08 mag) observed in ASASSN-25dc are also seen in some period bouncer systems, but not seen in systems well before the period bounce. Despite its short superhump period and indicated small mass ratio, we find no evidence of the excitement of the 2:1 tidal resonance. Moreover, its outburst rise timescale (\risets) is significantly longer than those measured at less than 0.4 d mag$^{-1}$ in other dwarf nova outbursts around the period minimum. Overall, an inside-out dwarf nova outburst from a massive disc in a system with a mass ratio around or even below the period minimum, but lacking the 2:1 tidal resonance, may explain all these observations. However, this challenges the existing models of dwarf nova superoutbursts, which do not predict these outburst properties in low-mass-ratio systems.

\end{abstract}


\clearpage
%%%%%%%%%%%%%%%%%%%%%%%%%%%%%%%%%%%%%%%%%%%%%%%%%%

%%%%%%%%%%%%%%%%% BODY OF PAPER %%%%%%%%%%%%%%%%%%

\section{Introduction}
\label{sec:1}

Cataclysmic variables (CVs) are close binary systems that host an accreting white dwarf (WD) and a mass-transferring low-mass secondary star that fills its Roche lobe \citep[see][for a general review]{war95book,hel01book}. In a non-magnetic CV, the transferred gas forms an accretion disc around the WD.
Because of the mass transfer and the angular momentum loss, CVs evolve towards shorter orbital periods and smaller mass ratios \citep{rob83CVperiod, kin88binaryevolution}. Magnetic braking is responsible for the angular momentum loss for a system above the period gap, while gravitational wave radiation is considered to play the role below the period gap. 
At the point when the secondary mass reaches 0.06--0.07 $M_\odot$ \citep[][]{kni11CVdonor, mca19DNeclipse}, the secondary becomes degenerate, altering its mass-radius relation. After this point, a CV system evolves towards a longer orbital period, observed as the period minimum around 80 min or 0.056 d \citep{gan09SDSSCVs}. CVs that have passed through the period minimum are called a period bouncer system.

Dwarf novae (DNe) form a major subtype in non-magnetic CVs. They show repeating outbursts with an amplitude of 1--9 mag, a duration of a few days--a month, and a cycle of weeks to decades in optical. These outbursts are understood in the thermal instability model in an accretion disc \citep{osa96review, kim20thesis, ham20CVreview}. In quiescence, a disc grows its mass with a constant mass transfer rate from the secondary star. Once its surface density reaches a critical value, the disc transitions into its high state with a higher accretion rate, observed as an outburst.
All DN outbursts can be divided into two classes with respect to which disc radii the heating wave starts to propagate: inside-out and outside-in outbursts. Because the critical surface density has a $r^{1.05}$ dependence on the disc radius $r$ \citep{can88outburst}, outside-in outbursts are expected to have a shorter rise timescale $\tau_\text{rise}$ to the outburst maximum \citep{min85DNDI}. This has indeed been observed in the case of SS Cyg, which shows a bimodal distribution of the rise timescale \citep{can98sscyg}.

CVs around the period minimum, typically with a mass ratio $q$ lower than 0.1, form a WZ Sge-type DN subclass (\citet{kat15wzsge} for a review). Their outbursts are accompanied by two types of short-period modulations known as superhumps: early and ordinary superhumps. These outbursts, accompanied by superhumps, are called superoutbursts. Early superhumps are double-peaked modulations with a superhump period $P_\text{SH}$ almost equal to the orbital one $P_\text{orb}$, which are suggested to originate from the double-arm structure in the disc excited by the 2:1 tidal resonance \citep{lin79lowqdisk, osa02wzsgehump, uem12ESHrecon}. Ordinary superhumps follow with a superhump period a few percent longer than the orbital period. This is understood as a precession of the eccentric disc deformed by the 3:1 tidal resonance \citep{whi88tidal, hir90SHexcess}. \citet{Pdot} established three stages of ordinary superhumps (stages A, B, and C) based on the superhump period changes. The superhump period derivative ($P_\text{dot} = \dot{P_\text{SH}} / P_\text{SH}$) is often defined for stage-B ordinary superhumps. Because the 2:1 resonance suppresses the growth of the 3:1 resonance \citep{lub91SHa}, early superhumps are always observed before ordinary superhumps.  It has been found that some outbursts in WZ Sge-type DNe lack apparent early superhumps due to (A) a low inclination, although the 2:1 resonance is excited and a faster powerlaw outburst decline corresponding to the early superhump phase is observed \citep[e.g., GW Lib; ][]{can01wzsge, Pdot}, and (B) a less energetic outburst, where a disc does not reach the 2:1 resonance radius and the 2:1 resonance is not excited \citep[e.g., the 2015 outburst of AL Com; ][]{kim16alcom}.

The behaviour of superhumps strongly depends on the mass ratio of the system. Indeed, several methods have been proposed to estimate the binary parameters from superhumps \citep[e.g.,][]{pat05SH, kat13qfromstageA}. \citet{nak14j0754j2304, kim18asassn16dt16hg} suggested some smoking-gun features for a DN outburst in a system with a mass ratio below the period minimum; (1) multiple rebrightening outbursts or a dip in brightness during the superoutburst plateau (double superoutburst), (2) long duration ($\geq$100 cycles) of stage-A superhump phase, (3) large period decrease ($\geq2\%$) between the stage-A and stage-B superhumps, (4) small superhump amplitude ($\leq0.1$ mag), (5) long delay ($\geq 10$ d) of ordinary superhump emergence, (6) a long outburst decline timescale ($\geq20$ d mag$^{-1}$) in the ordinary superhump phase, and (7) large outburst amplitude at the time of appearance of ordinary superhumps. We note that, as discussed in the above papers, not all period bouncers necessarily show all these features.

Some aspects of the outburst mechanism in WZ Sge-type DNe remain unclear \citep[e.g., see discussion in ][]{kat15wzsge, ham20CVreview}, especially on how their outbursts are triggered. 
Both the thermal-tidal instability and enhanced mass transfer models predict that superoutbursts in a system around the period minimum are always triggered as an outside-in outburst to account for the observed outburst energetics and cycles \citep{osa95wzsge, las95wzsge, ham97wzsgemodel, mey98wzsge}. The only observational evidence is their short rise timescales ($\simeq 0.2$ d mag$^{-1}$) observed in several samples with limited cadence \citep{pat81wzsge, mae09v455andproc, vic11gwlib}, and recently confirmed with TESS observations \citep{tam25wzsgetess}.

ASAS-SN reported the discovery of ASASSN-25dc on 2025 July 13.1 UTC (BJD 2460869.6) at $g=$13.07 mag. Both ASASSN and ATLAS detected the outburst rise since BJD 2460860.5, which attributes a slow rise to the outburst maximum for a week. The quiescence counterpart is the Gaia object 6129070308813533568 of $G=21.00(2)$ mag \citep{gaiaedr3}. Hence, the outburst amplitude is $\simeq$8.0 mag. There is no available parallax in Gaia. 
Ordinary superhumps were detected, and their preliminary period was determined as 0.05991(4) d, suggesting its classification as a WZ Sge-type DN (vsnet-alert 28089\footnote{\url{http://ooruri.kusastro.kyoto-u.ac.jp/mailarchive/vsnet-alert/28089}}).
This paper is structured as follows: Section \ref{sec:2} summarises an overview of our observations and analysis. Section \ref{sec:3} presents the result of the overall superoutburst and superhumps of ASASSN-25dc. In section \ref{sec:5}, we discuss our identification of ASASSN-25dc as an inside-out outburst in a period bouncer system, and further implications on the disc instability model. Section \ref{sec:6} summarises the paper. All observation epochs in this paper are described in the Barycentric Julian Day (BJD).

\section{Observations and Analysis}
\label{sec:2}

\subsection{Ground-based time-resolved observations}
Our ground-based time-resolved photometric observations of ASASSN-25dc were carried out through the Variable Star Network (VSNET) collaboration \citep{VSNET}. We also obtained the time-series observations using the Mookodi \citep{era24mookodi} on the Lesedi telescope and the SHOC \citep{cop13shoccam} on the 1.0-m telescope, both located at the Sutherland Observatory of the South African Astronomical Observatory. Their instrument details and observation logs are summarised in tables \ref{tab:A2} and \ref{tab:A3}, respectively. The zero point of the observations with the clear band was adjusted to the $V$-band magnitude of standard stars ($CV$ band).
Before period analysis, the global trend of the outburst decline was removed by subtracting a smoothed light curve obtained by locally weighted polynomial regression \citep[LOWESS; ][]{LOWESS}. The superhump maxima were determined following \citet{Pdot, kat13j1939v585lyrv516lyr}. The phase dispersion minimisation \citep[PDM; ][]{PDM} method was applied for the determination of superhump periods in this paper.  The 1$\sigma$ error for the PDM analysis is determined following \citet{fer89error, pdot2}. We list the epoch of the superhump maxima in Table \ref{tab:A4}.
 
\subsection{Archival time-domain surveys}

We also extracted the all-sky photometric survey data from the All-Sky Automated Survey for SuperNovae (ASAS-SN) Sky Patrol \citep{ASASSN, koc17ASASSNLC, ASASSNV2}, the Asteroid Terrestrial-impact Last Alert System \citep[ATLAS; ][]{ATLAS, hei18atlas, smi20atlas, shi21atlas}, {and the Catalina Surveys
Data Release 3 \citep[CRTS; ][]{CRTS}} to examine the global light curve profile in and before the outburst.

\section{Results}
\label{sec:3}

\begin{figure*}
 \begin{center}
  \includegraphics[width=\linewidth]{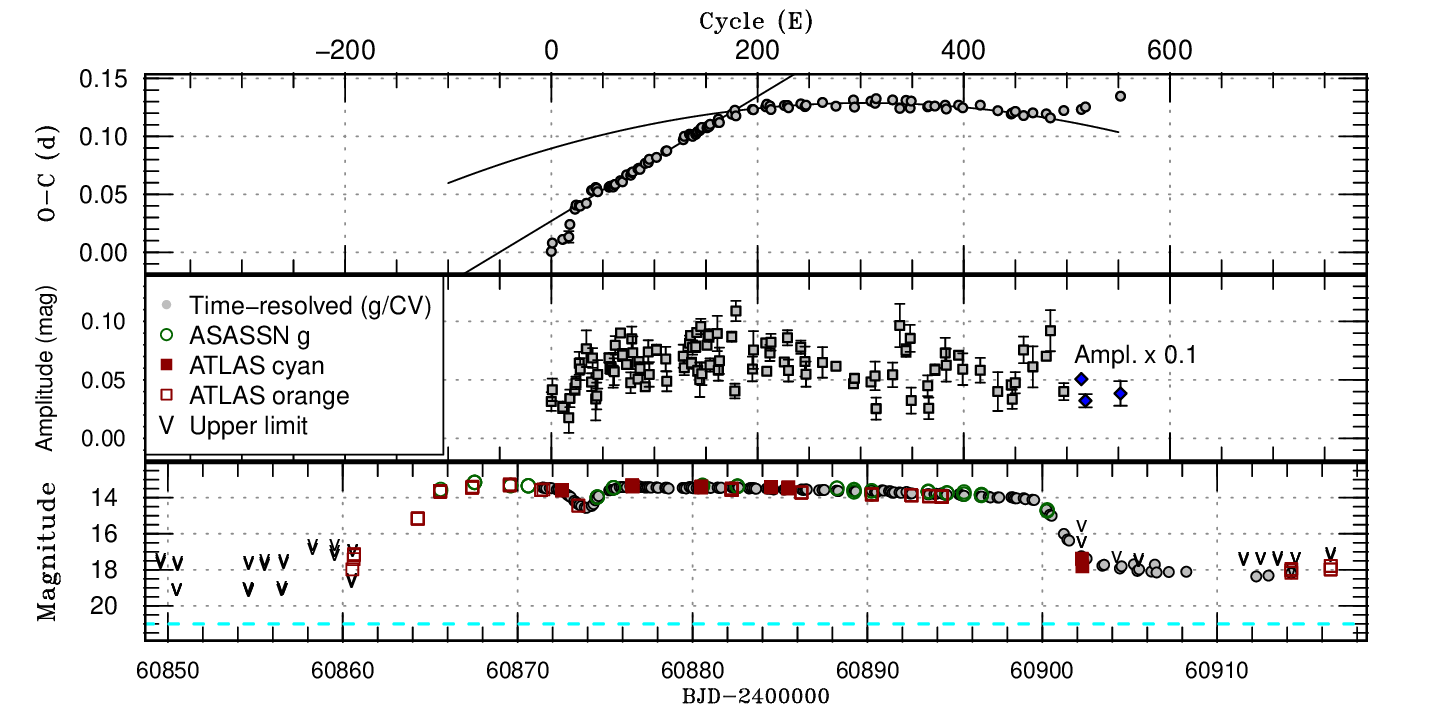}
 \end{center}
 \caption{
    Top panel; $O-C$ diagram of superhump maxima. $C= {\rm BJD~} 2460871.87867 + 0.05885 E$. The solid straight lines correspond to the periods of the stage-A superhumps. The curved line indicates the parabolic fit to the $O-C$ during the stage-B superhumps.
    Middle panel; Amplitude of superhumps in magnitude scale. The amplitudes after the rapid decline (after BJD 2460902.2; blue diamonds) are multiplied by 0.1 for visualisation purposes.
    Bottom panel; Overall optical light curve in outburst. The grey dots represent the data from VSNET and SAAO, binned in 0.1 d. The open circles, filled squares, and open squares represent the data of ASASSN $g$ band, ATLAS $c$ band, and ATLAS $o$ band, respectively. The V-shaped markers indicate the upper limits. The horizontal dashed line represents the brightness of the counterpart in the Gaia EDR3 ($G$=21.00(2) mag).
    }
\label{fig:longterm}
\end{figure*}

\begin{figure}
 \begin{center}
  \includegraphics[width=\linewidth]{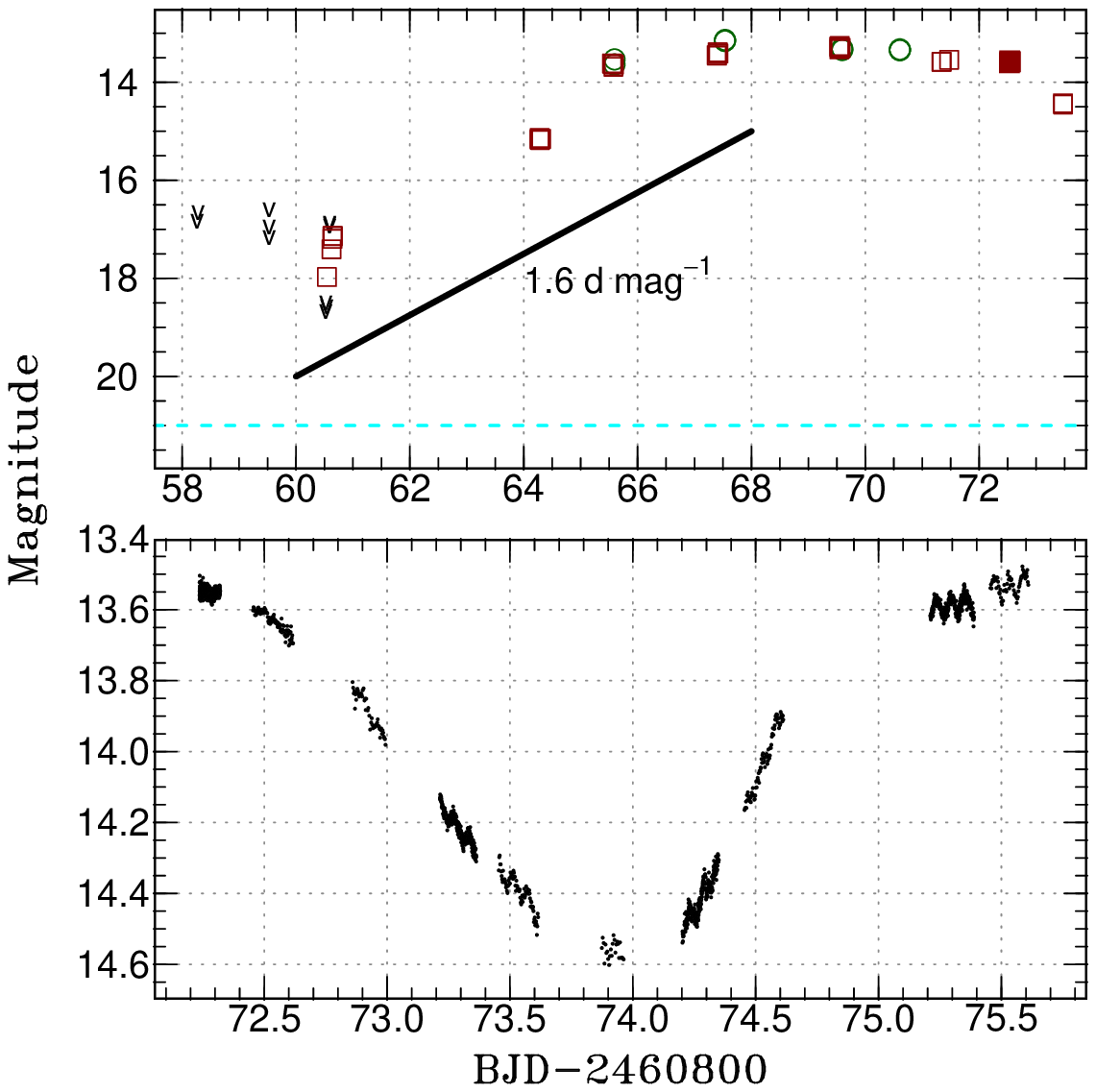}
 \end{center}
 \caption{
    Zoomed light curves around the outburst rise (top) and the dip (bottom). Symbols are the same as Figure \ref{fig:longterm}. The typical errors are smaller than the marker size. The solid line in the top panel represents the rise timescale of 1.6 d mag$^{-1}$.
    }
 \label{fig:zoomlc}
\end{figure}

\begin{figure*}
 \begin{center}
  \includegraphics[width=\linewidth]{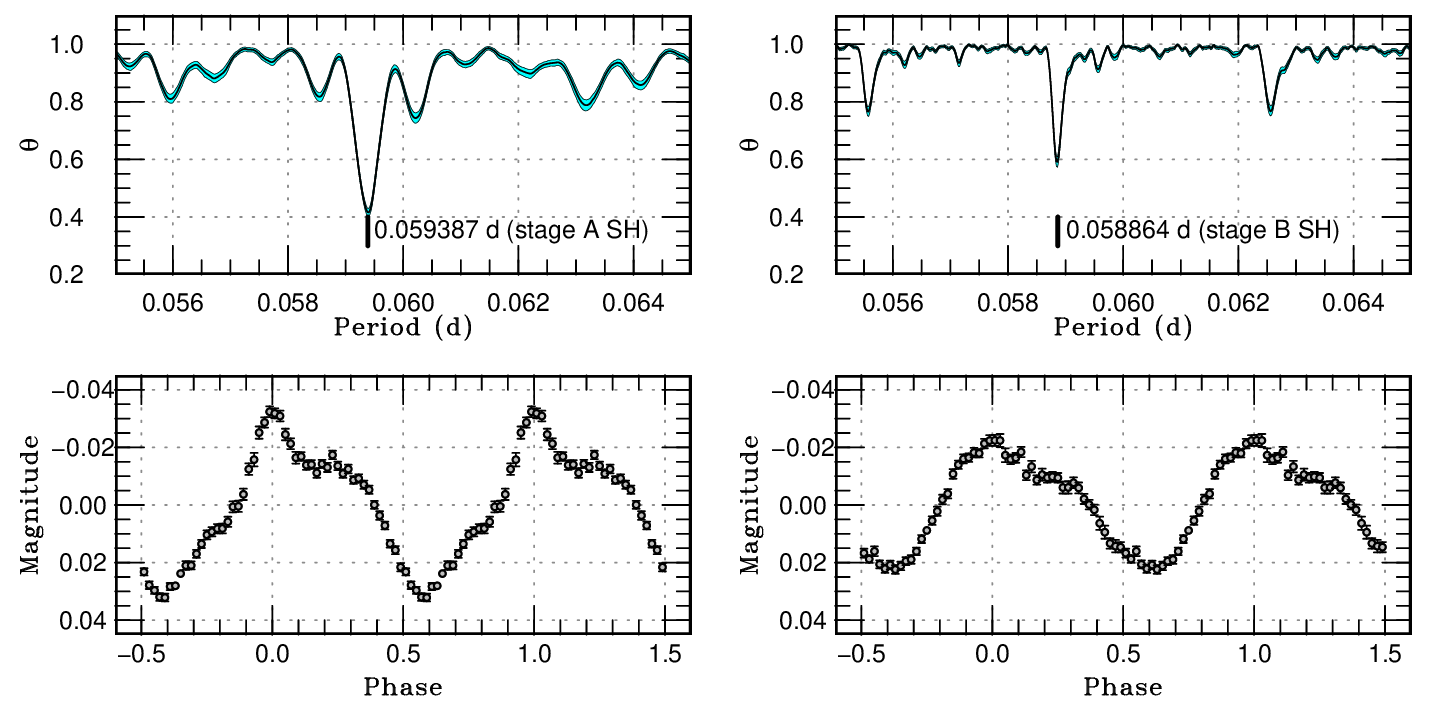}
 \end{center}
 \caption{
    Results of PDM analysis (top) and phase-folded profiles (bottom) of stage-A (left) and stage-B (right) superhumps. The superhump phase is arbitrary. The obtained periods are \stageAsh and \stageBsh d, respectively.
    }
 \label{fig:pdm}
\end{figure*}

\begin{figure}
 \begin{center}
  \includegraphics[width=\linewidth]{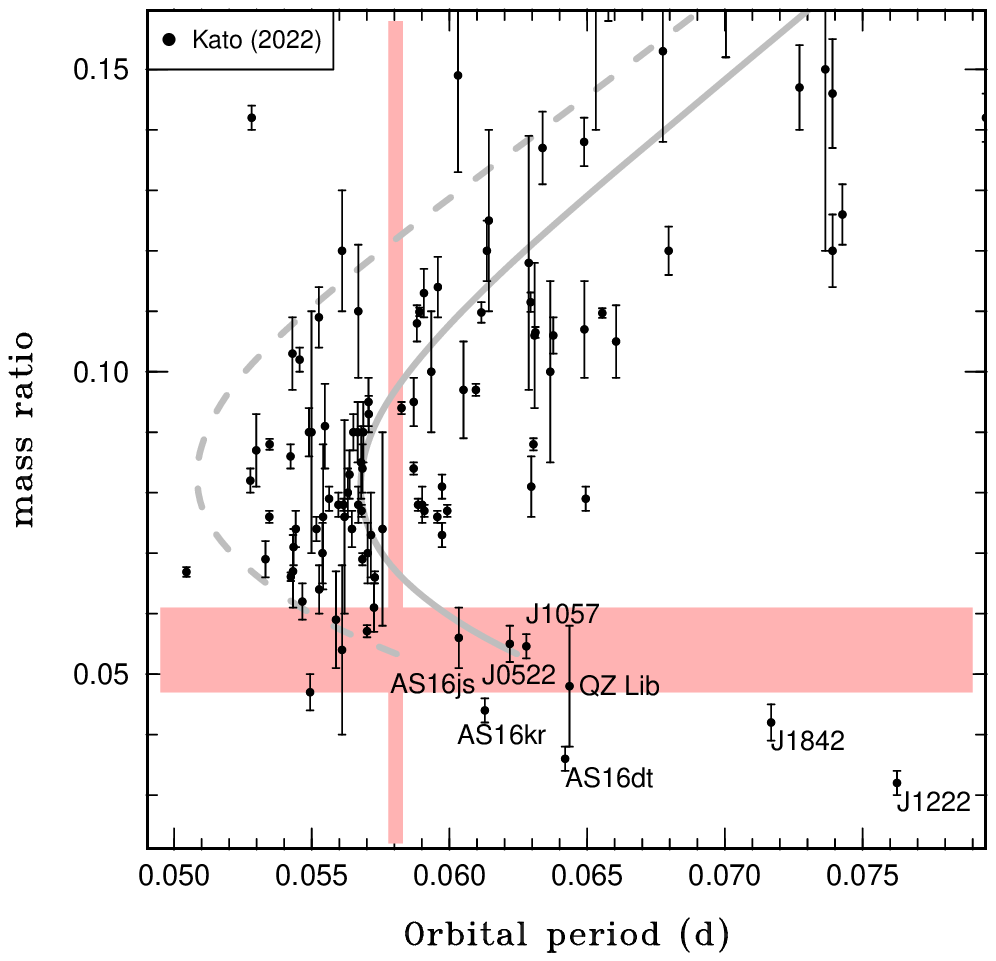}
 \end{center}
 \caption{
    Orbital periods versus mass ratios of CVs around the period minimum. The shaded region indicates the possible ranges of those of ASASSN-25dc. Black points represent the samples in \citet{kat22updatedSHAmethod}. The solid and dashed lines represent the standard and optimal evolutionary path of CVs in \citet{kni11CVdonor}. Some notable systems beyond the period minimum ($P_\text{orb} > 0.06$ d and $q < 0.06$) are labelled with text.
    Their abbreviations are as follows; AS16dt = ASASSN-16dt, AS16js = ASASSN-16js, AS16kr = ASASSN-16kr, J0522 = CRTS J052209.7$-$350530, J1057 = SDSS J105754.25$+$275947.5, J1222 = SSS J122221.7$-$311523, J1842 = PNV J18422792$+$4837425 = V529 Dra.
    }
 \label{fig:evolve}
\end{figure}

\subsection{Overall light curve}
\label{sec:3.1}

The bottom panel of Figure \ref{fig:longterm} presents the global light curve of ASASSN-25dc around its 2025 outburst. The outburst rise was first detected by ATLAS in its $o$ band on BJD 2460860.6 at $o\simeq$17.2 mag (top panel of Figure \ref{fig:zoomlc}). The light curve showed a slow rise to the maximum on BJD 2460869.5 at $o\simeq$13.4 mag. A linear interpolation using the ATLAS $o$-band data between BJD 2460860--2460868 yields its rise timescale as \risets. Even with the data BJD 2460864--2460868, the rise timescale is 1.5(2) d mag$^{-1}$. The first outburst plateau on BJD 2460869.5--2460872.0 is characterised by a flat-top and symmetric profile rather than a powerlaw outburst decline typically observed in WZ Sge-type DNe with the 2:1 tidal resonance.

After the first outburst plateau, ASASSN-25dc showed a V-shaped 1.0-mag dip for $\simeq$3.0 d on BJD 2460872.5--2460875.5 (bottom panel of Figure \ref{fig:zoomlc}). After recovering from the dip, the second outburst plateau reached $c\simeq13.3$ mag on BJD 2460876.5. The decline rate on the second plateau was \plateuts via a linear interpolation. The rapid decline from the outburst was observed on BJD 2460900.2.
Thus, the duration of the second plateau is $\simeq$25.0 d. The total duration of the outburst from its outburst rise is $\simeq$40 d, including the dip. These numbers are comparable to those of other WZ Sge-type DNe \citep[e.g., see Figure 11 of ][]{tam24j0302}.
After the rapid decline, no rebrightening outbursts have been observed, although a short rebrightening less than a day cannot be fully excluded due to the observation gaps.

ATLAS and ASASSN have observed ASASSN-25dc before this 2025 outburst since BJD 2457405 and 2456792, respectively. ATLAS provides continuous detections or upper limits around the quiescence level.  Although ASASSN reports some sparse detections, these points cannot be considered as real events because there are upper limits or detections at the quiescence level within the same or a few nights by ASASSN or ATLAS. {CRTS has 72 detections of the nearby star, Gaia object 6129070304518741376 ($G$=18.678(3) mag, 4.3$"$ from ASASSN-25dc), between BJD 2453589--2456047, while ASASSN-25dc has not been detected in this period, consistent with its quiescence level. Although we cannot exclude the possibility that ASASSN-25dc had an outburst in the observation gap of CRTS and ASASSN,} we conclude that there was no outburst of ASASSN-25dc at least in the past $\simeq$10 years, and presumably in the past $\simeq$20 years.

\subsection{Superhumps and binary parameters}
\label{sec:3.2}

The single-peaked ordinary superhumps are clearly present in the light curve during the second outburst plateau. Based on the $O-C$ diagram (the upper panel of Figure \ref{fig:longterm}), we determined the stage-A and stage-B superhump phases as BJD 2460873.3--2460881.6 and 2460882.5--2460899.5, respectively. The PDM analysis yields their superhump periods as \stageAsh and \stageBsh d, respectively (Figure \ref{fig:pdm}). The period decrease from stage-A to stage-B superhumps is 0.88(1)\%. 
The mean superhump amplitudes of stage-A and stage-B superhumps are 0.075 and 0.045 mag, respectively, and amplitudes of individual superhumps remain smaller than $\simeq0.1$ mag throughout the outburst (the middle panel of Figure \ref{fig:longterm}). The $P_\text{dot}$ during the stage-B superhumps was measured as \pdot cycle$^{-1}$.
As the system enters the rapid decline from the outburst, superhumps shifted to post-superoutburst superhumps with a longer period, judging from the $O-C$ diagram, although PDM analysis did not show any significant periodicity due to the low signal-to-noise ratio of the data. 

Judging in the $O-C$ diagram, the first identified peak of stage-A superhumps is on BJD 2460873.27 or $E=18$ before the bottom of the dip (bottom panel of Figure \ref{fig:zoomlc}). Some modulations are presented before this epoch. However, their peaks do not agree with the extrapolation from the maxima of stage-A superhumps. Additionally, we did not detect any periodicity around the superhump period before this epoch. Thus, we do not find any evidence of early superhumps in the first plateau, although our short ($\simeq2$ d) baseline of time-resolved observations before the dip does not allow us to completely exclude periodic variations before this epoch.

Because we did not detect any periods associated with the orbital period, we cannot use the relations between the mass ratio and the superhump period excess. However, Equation 30 in \citet{kat22updatedSHAmethod}, which formulates the empirical relation between the $P_\text{dot}$ during stage-B superhumps and mass ratio, can still be applied to ASASSN-25dc. Our $P_\text{dot}$ indicates its mass ratio as \massratio. We stress that, WZ Sge-type DNe with similar superhump periods but above the period minimum ($q\simeq0.1$) show positive $P_\text{dot}$ of $\sim 10 \times 10^{-5}$ cycle$^{-1}$ \citep{kat22updatedSHAmethod}, well larger than that of ASASSN-25dc.
Given this mass ratio, the expected fractional excess $\epsilon^*$ ($=1-P_\text{orb} / P_\text{SH}$) of the stage-A superhump period is 0.020(3) \citep[Table 1 in ][]{kat22updatedSHAmethod}, indicating its orbital period as 0.0580(3) d. We note that future observations to determine its orbital period are still vital to confirm these binary parameters, including the mass ratio.

\section{Discussion}
\label{sec:5}

\subsection{ASASSN-25dc as a period bouncer candidate}

Our empirical orbital period and mass ratio of ASASSN-25dc indicate that ASASSN-25dc is located around the period minimum, and is even a candidate period bouncer (Figure \ref{fig:evolve}).
Other outburst characteristics of ASASSN-25dc also agree with those observed in some period bouncers and candidates \citep{kim18asassn16dt16hg}: the long duration of stage-A superhumps ($\simeq$150 cycles), small superhump amplitude ($\simeq0.08$ mag), a long decline timescale of the outburst plateau (\plateuts), and a large outburst amplitude at the time of emergence of ordinary superhumps (7.7 mag after the dip). On the other hand, ASASSN-25dc does not exhibit a large decrease between the stage-A and stage-B superhump periods. However, this may not be so critical, because the confirmed period bouncer SSS J122221.7$-$311523 and the suggested one ASASSN-16dt also show a small period decrease of an order of $\leq$1\% \citep{kim18asassn16dt16hg}.

Despite its short superhump period and proposed small mass ratio, we did not detect early superhumps in ASASSN-25dc. Although the amplitude of early superhumps and hence its detectability depend on the inclination of a system, a faster powerlaw decline in the early superhump phase is universally observed, including a low-inclination system \citep{kat15wzsge} and a system with the double-superoutburst property \citep{nak14j0754j2304, kim18asassn16dt16hg}. This clearly differs from the flat-top profile of the first outburst plateau in ASASSN-25dc. A similar flat-top profile of the first outburst plateau was observed in SSS J122221.7$-$311523 \citep{kat13j1222}. This system also showed a long stage-A superhump phase ($\simeq$200 cycles) and slow decline rate ($\simeq$50 d mag$^{-1}$), although it has the even longer orbital period (0.07625(5) d) and lower mass ratio (0.032(2)) compared to ASASSN-25dc \citep{kat13j1222, neu17j1222, kim18asassn16dt16hg}. We here conclude that the 2:1 resonance indeed was not excited in ASASSN-25dc, rather than early superhumps were missed due to an observational bias of either a lower inclination or a short baseline of our time-resolved observations.

\subsection{A slow-rise outburst in ASASSN-25dc}

One of the significant differences of ASASSN-25dc from WZ Sge-type DNe is its long rise timescale. Table \ref{tab:1} compares the rise timescales of ASASSN-25dc and other WZ Sge-type DNe. It must be noted that the samples in \citet{otu16DNstats} should be treated as an upper limit, as their rise timescales are determined solely by CRTS with 1-d cadence. This clearly demonstrates that the observed rise timescale in ASASSN-25dc is unambiguously longer than those less than $0.4$ d mag$^{-1}$ in WZ Sge-type DNe.
In the case of SS Cyg, \citet{can98sscyg} showed that the rise timescale shows a bimodal distribution, of which the faster timescales are expressed as 0.56$\pm0.14$ d mag$^{-1}$ while the slower timescales exceed 1.0 d mag$^{-1}$. 
In Table \ref{tab:1}, we also measure the rise timescales of outside-in and inside-out outbursts of an SU UMa-type DN V516 Lyr using Kepler data \citep[][; see also Appendix \ref{sec:A1}]{kat13j1939v585lyrv516lyr}. This also shows a similar distribution of rise timescales of outside-in and inside-out outbursts to those of SS Cyg.
We note that the slightly longer rise timescales of outside-in outbursts in SS Cyg and V516 Lyr, than those of WZ Sge-type DNe, are likely due to their brighter quiescence. 
Thus, the observed rise timescale of ASASSN-25dc lies within the range of inside-out outbursts of SS Cyg and V516 Lyr, unlike other WZ Sge-type DNe.

\begin{table}
	\centering
    \caption{The outburst rise timescales of WZ Sge-type DNe and V516 Lyr}\label{tab:1}
	\begin{tabular}{ccc} 
  \hline              
    Name & $\tau_\text{rise}$ &  Reference\commenta\\ 
     &  d mag$^{-1}$  & \\ 
    \hline
    ASASSN-25dc & 1.62(9)  & this work \\
    ASASSN-23ba & 0.208(3) & [1] \\ 
    PNV J1903$-$31 & 0.169(1) & [1]  \\ 
    V748 Hya & 0.066(1) & [1]  \\ 
    ASASSN-25ci & 0.317(3) & [1]  \\ 

    V455 And & 0.1 & [2]  \\    
    GW Lib & 0.1 & [3]  \\
    MASTER J0302$+19$ & 0.31(1) & [4]  \\

    AL Com & 0.2 & [5]  \\
    UZ Boo & 0.3 & [5] \\
    V1108 Her & 0.3 & [5] \\

    \hline
    V516 Lyr (outside-in) & 0.55(6)  & [6] \\
    V516 Lyr (inside-out) & 1.15(7)  & [6] \\
    \hline
    \multicolumn{3}{l}{\commenta [1]\citet{tam25wzsgetess}, [2]\citet{mae09v455andproc},}\\ 
    \multicolumn{3}{l}{[3]\citet{vic11gwlib}, [4]\citet{tam24j0302},  }\\
    \multicolumn{3}{l}{[5]\citet{otu16DNstats}, [6]\citet{kat13j1939v585lyrv516lyr}.}\\
	\end{tabular}
\end{table}

\subsection{Implications in the disc instability model}

A dip in the outburst plateau separating the early and ordinary superhump phases has been regarded as a smoking-gun feature of period-bouncer systems \citep[e.g. ][and references therein]{kim18asassn16dt16hg}. Their dip is understood as follows: The first superoutburst is maintained by the 2:1 resonance. Since the growth of the 3:1 resonance takes a longer time in a low-mass ratio system and is also suppressed by the 2:1 resonance \citep[][]{lub91SHa, lub91SHb}, the cooling wave started to propagate before the 3:1 tidal instability gets fully excited, which is observed as the growing ordinary superhumps in the dip. As the 3:1 resonance fully sets in, the disc returns to its outburst state, resulting in the typical behaviour of ordinary superhumps in the second superoutburst. 

In the case of ASASSN-25dc, we consider that, without the excitement of the 2:1 resonance, the first outburst plateau is a precursor outburst powered by thermal instability and viscous depletion in a massive \citep[$\simeq10^{24}$ g in a case of WZ Sge compared to $\simeq10^{23}$ g in standard SU UMa stars;][]{ich93SHmasstransferburst, sma93wzsge} disc \citep[type B outburst in][]{osa05DImodel}, rather than a superoutburst powered by the 2:1 tidal resonance. A similar scenario is suggested in SSS J122221.7$-$311523 \citep{kat13j1222} and in EG Cnc \citep{kim21EGCnc}. This naturally explains the relatively short duration of the first outburst plateau compared to that of its second outburst plateau with ordinary superhumps. Once the 3:1 tidal resonance gets fully excited, the disc undergoes a superoutburst. Its just $\simeq$4-d delay of the emergence of ordinary superhumps from the outburst maximum, shorter than those of other low-mass ratio systems \citep{kim18asassn16dt16hg}, would be understood as the lack of the suppression of the 3:1 resonance by the 2:1 resonance, although it is unclear from the light curve exactly when the disc reached the 3:1 resonance radius.

The superoutburst of ASASSN-25dc is another example of a superoutburst from a low ($q\leq0.1$) mass-ratio system, but lacking evidence of the excitation of the 2:1 tidal resonance. 
It has been found that some genuine WZ Sge-type DNe undergo a superoutburst lacking an early superhump phase, within a few years after the more energetic superoutburst with an early superhump phase. \citet{kim16alcom, kim21EGCnc} discussed that an outburst somehow triggered in a less massive disc may explain this phenomenon. Such a scenario, however, cannot be applied to ASASSN-25dc, considering its long ($\simeq$40 d) outburst duration and no previous outbursts for at least the last 10 years before this 2025 superoutburst.
Moreover, the 2022 superoutburst of the WZ Sge-type DN TCP J05515391+6504346, which lacked the early superhump phase, showed the rise timescale of 0.070(4) d mag$^{-1}$, within the range of other WZ Sge-type DNe with the early superhump phase \citep{tam25wzsgetess}. Thus, a less massive disc does not necessarily lead to a longer rise timescale.

An alternative scenario we propose here is an inside-out superoutburst in the low-mass-ratio system ASASSN-25dc. 
The rise timescale of DN outbursts is regarded as a strong indication of whether an outburst is triggered as an outside-in or inside-out type outburst. Its long timescale within the range of inside-out outbursts of SS Cyg and V516 Lyr strongly supports this scenario. The symmetric profile of the first outburst plateau also agrees with this interpretation. 
\citet{sma84DNoutburst, min85DNDI} show that indeed a disc remains smaller in inside-out outbursts than in outside-in outbursts. This effect potentially prevents its disc from reaching the 2:1 resonance radius and exciting the tidal instability in the first outburst plateau, even with its mass ratio low enough to excite the 2:1 tidal resonance.

This scenario, an inside-out outburst in a system around the period minimum, challenges the existing models explaining outbursts in WZ Sge-type DNe. 
Outbursts with a slow rise, suggesting an inside-out outburst, have been observed in a wide range of orbital periods from standard SU UMa-type DNe \citep[e.g., ][]{kat13j1939v585lyrv516lyr, skl18nyser} to long-orbital-period DNe \citep[][]{kim92gkper}.
However, various authors have found that simply lowering the mass transfer rate in a model for standard SU UMa-type DNe results in inside-out outbursts but with much less accreted mass \citep{sma93wzsge, las95wzsge, osa95wzsge}, inconsistent with the long outburst duration of ASASSN-25dc and other WZ Sge-type DNe. To take into account the outburst energetics of WZ Sge-type DNe, two models have been mainly proposed: an inner disc truncation or an extremely low quiescence viscosity.
In the case of an inner disc truncation and moderate quiescence viscosity ($\alpha_\text{c}\simeq0.01$), \citet{las95wzsge, ham97wzsgemodel} have suggested that an outburst must be triggered and maintained by the enhanced mass transfer from the secondary star since the disc mass before the outburst is much less than the critical surface density. This must always result in an outside-in outburst regardless of the state of the disc before the outburst. The simulated light curve in \citet{ham97wzsgemodel} indeed shows a rapid rise to the outburst maximum, unlike that of ASASSN-25dc.
\citet{osa95wzsge, mey98wzsge} have found that a model with an extremely low quiescence viscosity ($\alpha_\text{c} \leq 0.001$) still requires a massive and large disc at the onset of an outburst, as the critical surface density is larger in the outer disc. The simulated light curve in \citet{osa95wzsge} also shows a rapid rise to the outburst maximum.
One possible solution for altering this to an inside-out outburst is radial-dependent viscosity in quiescence. Larger viscosity $\alpha$ in an inner disc allows a shorter viscosity drift timescale and more mass in the inner disc. Some simulation studies found its radial dependence to be $\alpha \propto r^{0.5}$  \citep{min89quiescenceviscosity, can93DI}. Although a flatter radial dependence in the quiescent viscosity compared to other WZ Sge-type DNe may enable a massive disc to trigger an inside-out outburst, this still must require a rather different disc and system structure from other systems. 
Future observations, speculatively in the respects of the WD magnetosphere or the evaporation effect in the inner disc, and numerical simulations, to find a possible range for inner disc truncation and radial dependence of viscosity that can reproduce the light curve of ASASSN-25dc, are necessary to distinguish these scenarios.

\section{Summary}
\label{sec:6}

We have reported the optical time-resolved observations of a new dwarf nova system, ASASSN-25dc, during its 2025 superoutburst. We find superhumps in the outburst plateau with the periods of \stageAsh and \stageBsh d for stage-A and stage-B ordinary superhumps, respectively. The negative $P_\text{dot}$ in stage-B superhumps (\pdot cycle$^{-1}$) indicates its mass ratio as \massratio. The slow decline rate of the ordinary superhump phase and small superhump amplitude also support that ASASSN-25dc is a CV system around or even after the period bounce. The flat-top and symmetric profile of the first outburst plateau proposes that the 2:1 tidal resonance was not excited. Furthermore, its initial outburst rise is characterised by a rise timescale of \risets, significantly longer than those of other well-studied WZ Sge-type DNe. This suggests that the 2025 superoutburst of ASASSN-25dc was initially triggered as an inside-out outburst.
These points illustrate a scenario in which ASASSN-25dc is the first low-mass-ratio system with an inside-out dwarf nova superoutburst. This fact itself challenges the existing disc instability models explaining outbursts in WZ Sge-type DNe. An inner disc truncation model with an outburst triggered by the enhanced mass transfer cannot reproduce an inside-out outburst. A significant modification, especially in the disc and system structure in quiescence, is still necessary to explain all the observed phenomena in the framework of the thermal-tidal instability model with very low viscosity.

\section*{Acknowledgements}

We acknowledge amateur and professional astronomers around the world who have shared data on variable stars and transients with the VSNET collaboration. This paper uses observations made from the South African Astronomical Observatory (SAAO).
This work has made use of data from the Asteroid Terrestrial-impact Last Alert System (ATLAS) project. The Asteroid Terrestrial-impact Last Alert System (ATLAS) project is primarily funded to search for near earth asteroids through NASA grants NN12AR55G, 80NSSC18K0284, and 80NSSC18K1575; byproducts of the NEO search include images and catalogs from the survey area. The ATLAS science products have been made possible through the contributions of the University of Hawaii Institute for Astronomy, the Queen’s University Belfast, the Space Telescope Science Institute, the South African Astronomical Observatory, and The Millennium Institute of Astrophysics (MAS), Chile.

%%%%%%%%%%%%%%%%%%%%%%%%%%%%%%%%%%%%%%%%%%%%%%%%%%
\section*{Data Availability}

Most of the ground-based observations are publicly available at the American Association of Variable Star Observers (AAVSO) International Database (\url{https://www.aavso.org/}). Other data is available upon request to individual observers.



\bibliographystyle{mnras}
\bibliography{cvs} % if your bibtex file is called example.bib

%\begin{thebibliography}{99}
%\end{thebibliography}

%%%%%%%%%%%%%%%%%%%%%%%%%%%%%%%%%%%%%%%%%%%%%%%%%%

%%%%%%%%%%%%%%%%% APPENDICES %%%%%%%%%%%%%%%%%%%%%
\clearpage
\appendix
% If you want to present additional material which would interrupt the flow of the main paper,
% it can be placed in an Appendix which appears after the list of references.

\renewcommand{\thetable}{A\arabic{table}}
\renewcommand{\thefigure}{A\arabic{figure}}
\newcolumntype{V}{!{\color{black}\vrule width 0.6pt}}

\section{Rise timescales of V516 Lyr}
\label{sec:A1}

\begin{figure}
 \begin{center}
  \includegraphics[width=\linewidth]{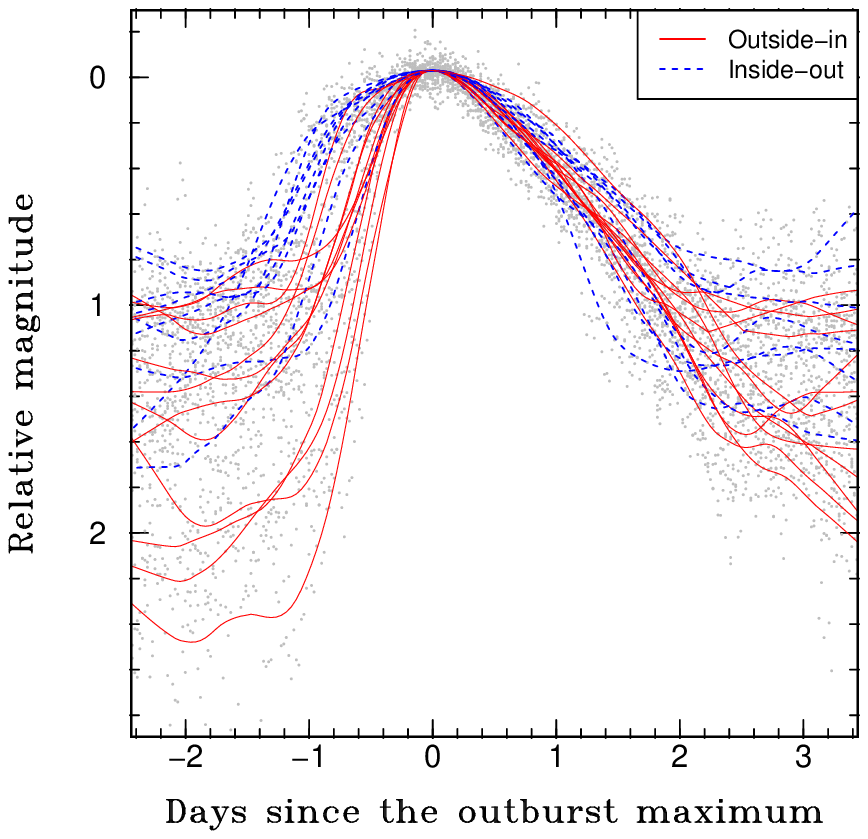}
 \end{center}
 \caption{
    The outburst light curves of V516 Lyr observed by Kepler (gray circles). They are normalized at the outburst maximum and its epoch. The blue-dashed and red-solid lines represent the smoothed ones of inside-out and outside-in outbursts.
    }
 \label{fig:v516}
\end{figure}

Although the rise timescale measurements of SS Cyg in \citet{can98sscyg} provide a meaningful sample, SS Cyg is much brighter in quiescence, and its disc is larger than CVs around the period minimum, which apparently affects the rise timescale. Moreover, they claim that their sample of SS Cyg still may suffer from the 1-d aliases.
To reliably measure the rise timescales of inside-out outbursts in a system with a short orbital period (i.e., below the period gap), we utilized the light curve of V516 Lyr ($P_\text{orb}$ = 0.083999(8) d) observed by Kepler with a 30-min cadence. We used only the outbursts confidently labelled as either a normal outside-in or inside-out outbursts in \citet{kat13j1939v585lyrv516lyr}. 
To qualitatively measure the rise timescale, we first obtained the smoothed light curve by LOWESS and determined the epoch of the outburst maximum (Figure \ref{fig:v516}). We then fitted the light curve in the magnitude scale with a broken linear function ($F(t)$; Equation \ref{eq:1}) using the data between $0.2$--$2.5$ days before the outburst maximum, covering both the quiescence and outburst rise,

\begin{equation}
\label{eq:1}
F(t) =
\begin{cases}
c, & t \le t_\text{outburst} \\
c - \frac{1}{\tau_\text{rise}}\,(t - t_\text{outburst}), & t > t_\text{outburst}
\end{cases}
\end{equation}

\noindent
where $t_\text{outburst}$, $\tau_\text{rise}$, and $c$ are the epoch when the outburst starts, the rise timescale, and the constant brightness in quiescence.

Table \ref{tab:A1} lists the measured rise timescales of 21 outbursts, in the unit of d mag$^{-1}$, 12 of which are outside-in outbursts, and the remaining 9 are inside-out outbursts. Among the outbursts listed as outside-in, Q10-3, Q10-9, and Q14-7 show a shoulder on the outburst rise, which prolongs the measured rise timescales. Although Q9-1 and Q9-4 are classified as inside-out outbursts, they show short rise timescales comparable to those of outside-in outbursts. Indeed, their outburst profile resembles some of the outside-in outbursts. As \citet{kat13j1939v585lyrv516lyr} claimed, since their outburst type is flagged by eye, we excluded these five outbursts, and measured the median rise timescales of 9 outside-in and 7 inside-out outbursts as the confident sample, which are listed in Table  \ref{tab:1}.

\section{Observation logs}

\begin{table}
	\centering
    \caption{Rise timescales of V516 Lyr observed by Kepler}\label{tab:A1}
	\begin{tabular}{cccc} 
  \hline              
    Outburst\commenta & Type & $\tau_\text{rise}$ & Error\\ 
     &  & d mag$^{-1}$ & d mag$^{-1}$ \\ 
    \hline
Q6-4 & outside-in &0.38&0.03\\
Q6-5 & outside-in &0.29&0.02\\
Q7-1 & outside-in &0.66&0.06\\
Q7-3 & outside-in &0.29&0.02\\
Q7-4 & outside-in &0.45&0.03\\
Q8-4 & outside-in &0.37&0.03\\
Q9-1 & inside-out &0.44&0.02\\
Q9-3 & inside-out &1.12&0.03\\
Q9-4 & inside-out &0.67&0.03\\
Q9-5 & outside-in &0.88&0.10\\
Q9-7 & inside-out &1.03&0.06\\
Q10-2 & inside-out &1.08&0.07\\
Q10-3\commentb & outside-in &0.76&0.05\\
Q10-7 & outside-in &0.54&0.06\\
Q10-8 & outside-in &0.55&0.03\\
Q10-9\commentb & outside-in &0.77&0.05\\
Q14-1 & inside-out &1.15&0.05\\
Q14-2 & inside-out &1.35&0.06\\
Q14-3 & inside-out &1.49&0.10\\
Q14-6 & inside-out &1.22&0.06\\
Q14-7\commentb & outside-in &0.73&0.03\\
    \hline
    \multicolumn{4}{l}{\commenta Outburst ID in \citet{kat13j1939v585lyrv516lyr}.}\\
    \multicolumn{4}{l}{\commentb Outburst with a shoulder rise.}\\
	\end{tabular}
\end{table}

\begin{table*}
	\centering
    \caption{List of instruments for photometric observation}\label{tab:A2}
	\begin{tabular}{ccc} 
  \hline              
    Observer's code & Telescope($\&$ CCD) & Observatory (or Observer)\\ 
    \hline
    GPX & 25cm reflector+ZWO ASI178mm & W. Goltz\\
    HaC & 35cm C14+QHY600M & F.~J. Hambsch \\
    LSD & 1.0m Lesedi+Mookodi & SAAO\\
    MLF & 30cm RCX400+ST8-XME & B. Monard\\
    S40 & 1.0m+SHOC & SAAO\\
    SPE & 30cm RCX400+ST8-XME & P. Starr\\
    \hline
	\end{tabular}
\end{table*}

\begin{table*}
	\centering
    \caption{Log of ground-based time-resolved photometric observations}\label{tab:A3}
	\begin{tabular}{ccccccc V ccccccc} 
  \hline              
    Start\commenta & End\commenta & Mag\commentb & $\sigma_{\rm Mag}$\commentc & $N$\commentd & Obs\commente & Band & 
    Start\commenta & End\commenta & Mag\commentb & $\sigma_{\rm Mag}$\commentc & $N$\commentd & Obs\commente & Band\\ 
    \hline
871.4547 & 871.6204 & 13.458 & 0.002 & 93 & HaC & CV & 888.2090 & 888.2952 & 13.358 & 0.001 & 411 & LSD & g \\
871.8624 & 871.9826 & 13.417 & 0.002 & 50 & GPX & CV & 889.2095 & 889.3050 & 13.390 & 0.001 & 468 & LSD & g \\
872.2368 & 872.3203 & 13.418 & 0.000 & 660 & LSD & g & 889.2122 & 889.3567 & 13.427 & 0.001 & 291 & MLF & CV \\
872.4545 & 872.6173 & 13.635 & 0.003 & 97 & HaC & CV & 890.2185 & 890.2595 & 13.437 & 0.001 & 230 & LSD & g \\
872.8595 & 872.9936 & 13.836 & 0.007 & 52 & GPX & CV & 890.4579 & 890.5668 & 13.689 & 0.003 & 66 & HaC & CV \\
873.2150 & 873.3630 & 14.001 & 0.002 & 295 & MLF & CV & 890.8684 & 890.9369 & 13.690 & 0.006 & 28 & GPX & CV \\
873.4546 & 873.6147 & 14.397 & 0.005 & 93 & HaC & CV & 891.2566 & 891.2976 & 13.486 & 0.001 & 230 & LSD & g \\
873.8740 & 873.9627 & 14.505 & 0.005 & 25 & GPX & CV & 891.4585 & 891.5116 & 13.715 & 0.003 & 29 & HaC & CV \\
874.2021 & 874.3497 & 14.200 & 0.003 & 289 & MLF & CV & 891.8834 & 891.9608 & 13.818 & 0.007 & 33 & GPX & CV \\
874.4548 & 874.6122 & 14.023 & 0.009 & 93 & HaC & CV & 892.2249 & 892.2656 & 13.523 & 0.002 & 193 & LSD & g \\
875.2111 & 875.3886 & 13.374 & 0.001 & 355 & MLF & CV & 892.4584 & 892.5616 & 13.760 & 0.003 & 65 & HaC & CV \\
875.4555 & 875.6098 & 13.529 & 0.003 & 79 & HaC & CV & 893.4584 & 893.5584 & 13.810 & 0.002 & 57 & HaC & CV \\
875.8575 & 876.0388 & 13.562 & 0.002 & 200 & SPE & CV & 893.8649 & 893.9446 & 13.967 & 0.002 & 100 & SPE & CV \\
876.1947 & 876.2813 & 13.369 & 0.001 & 1475 & S40 & g & 894.2093 & 894.2501 & 13.605 & 0.001 & 229 & LSD & g \\
876.4557 & 876.6063 & 13.371 & 0.003 & 69 & HaC & CV & 894.4585 & 894.5561 & 13.839 & 0.003 & 51 & HaC & CV \\
876.8560 & 877.0402 & 13.535 & 0.002 & 214 & SPE & CV & 895.2346 & 895.2932 & 13.660 & 0.001 & 329 & LSD & g \\
877.2573 & 877.3332 & 13.678 & 0.001 & 1172 & S40 & CV & 895.4590 & 895.5538 & 13.895 & 0.003 & 53 & HaC & CV \\
877.4553 & 877.6043 & 13.423 & 0.003 & 68 & HaC & CV & 896.4592 & 896.5495 & 13.940 & 0.003 & 46 & HaC & CV \\
877.9259 & 878.0296 & 13.590 & 0.002 & 131 & SPE & CV & 896.9560 & 897.0374 & 14.151 & 0.003 & 103 & SPE & CV \\
878.4554 & 878.6005 & 13.452 & 0.003 & 71 & HaC & CV & 897.4593 & 897.5477 & 13.986 & 0.003 & 46 & HaC & CV \\
879.4560 & 879.5969 & 13.466 & 0.003 & 65 & HaC & CV & 898.2200 & 898.3487 & 13.805 & 0.002 & 255 & MLF & CV \\
879.8565 & 880.0783 & 13.617 & 0.002 & 279 & SPE & CV & 898.2248 & 898.2835 & 13.815 & 0.001 & 328 & LSD & g \\
880.2102 & 880.3583 & 13.246 & 0.002 & 286 & MLF & CV & 898.4592 & 898.5452 & 14.034 & 0.002 & 44 & HaC & CV \\
880.2241 & 880.3217 & 13.220 & 0.001 & 335 & LSD & g & 898.9403 & 899.0234 & 14.233 & 0.003 & 105 & SPE & CV \\
880.4561 & 880.5953 & 13.445 & 0.004 & 64 & HaC & CV & 899.4640 & 899.5415 & 14.127 & 0.004 & 43 & HaC & CV \\
880.8546 & 881.0584 & 13.605 & 0.002 & 255 & SPE & CV & 900.2131 & 900.2802 & 14.493 & 0.001 & 335 & LSD & g \\
881.4561 & 881.5926 & 13.459 & 0.003 & 108 & HaC & CV & 900.4596 & 900.5398 & 14.989 & 0.006 & 45 & HaC & CV \\
882.2669 & 882.3307 & 13.195 & 0.001 & 340 & LSD & g & 901.2130 & 901.2806 & 15.849 & 0.002 & 337 & LSD & g \\
882.4562 & 882.5900 & 13.462 & 0.003 & 111 & HaC & CV & 901.4597 & 901.5365 & 16.365 & 0.010 & 41 & HaC & CV \\
883.2578 & 883.2988 & 13.227 & 0.002 & 230 & LSD & g & 902.2115 & 902.2702 & 17.720 & 0.009 & 321 & LSD & g \\
883.4562 & 883.5869 & 13.474 & 0.004 & 73 & HaC & CV & 902.4596 & 902.5343 & 17.417 & 0.019 & 51 & HaC & CV \\
884.2104 & 884.3361 & 13.278 & 0.002 & 252 & MLF & CV & 903.4599 & 903.5297 & 17.740 & 0.022 & 37 & HaC & CV \\
884.2359 & 884.3226 & 13.236 & 0.001 & 460 & LSD & g & 904.4601 & 904.5283 & 17.850 & 0.024 & 37 & HaC & CV \\
884.4568 & 884.5839 & 13.503 & 0.003 & 134 & HaC & CV & 905.2146 & 905.2150 & 18.161 & 0.049 & 3 & LSD & g \\
885.2755 & 885.3142 & 13.272 & 0.002 & 217 & LSD & g & 905.4601 & 905.5248 & 18.005 & 0.027 & 38 & HaC & CV \\
885.4568 & 885.5810 & 13.520 & 0.003 & 107 & HaC & CV & 906.2297 & 906.2302 & 18.568 & 0.124 & 3 & LSD & g \\
886.2293 & 886.2964 & 13.296 & 0.001 & 323 & LSD & g & 906.4605 & 906.5229 & 18.029 & 0.049 & 35 & HaC & CV \\
886.4574 & 886.5782 & 13.545 & 0.003 & 69 & HaC & CV & 907.2526 & 907.2530 & 18.588 & 0.055 & 3 & LSD & g \\
887.2313 & 887.2340 & 13.336 & 0.002 & 16 & LSD & g & 908.2207 & 908.2211 & 18.573 & 0.061 & 3 & LSD & g \\
887.4575 & 887.5761 & 13.572 & 0.005 & 71 & HaC & CV & 912.2438 & 912.2443 & 18.835 & 0.086 & 3 & LSD & g \\    \hline
    \multicolumn{14}{l}{\commenta BJD$-$2460000}\\
    \multicolumn{14}{l}{\commentb Mean magnitude.} \\
    \multicolumn{14}{l}{\commentc Standard deviation of the observed magnitude.} \\
    \multicolumn{14}{l}{\commentd Number of observations.} \\
    \multicolumn{14}{l}{\commente Observer's codes are same as table \ref{tab:1} }
	\end{tabular}
\end{table*}

\begin{table*}
	\centering
    \caption{Times of superhump maxima}\label{tab:A4}
	\begin{tabular}{ccccc V ccccc} 
  \hline              
    $E$ & maximum time\commenta & error & $O-C$\commentb & $N$\commentc & $E$ & maximum time\commenta & error & $O-C$\commentb & $N$\commentc\\ 
    \hline
0 & 871.87943 & 0.00137 & 0.00076 & 18 & 154 & 881.05201 & 0.00082 & 0.11044 & 38 \\
1 & 871.94550 & 0.00165 & 0.00798 & 19 & 161 & 881.46573 & 0.00063 & 0.11221 & 24 \\
11 & 872.53709 & 0.00122 & 0.01107 & 24 & 162 & 881.52726 & 0.00099 & 0.11489 & 32 \\
17 & 872.89245 & 0.00496 & 0.01333 & 18 & 163 & 881.58314 & 0.00153 & 0.11192 & 27 \\
18 & 872.96201 & 0.00163 & 0.02404 & 19 & 175 & 882.29658 & 0.00018 & 0.11916 & 249 \\
23 & 873.26943 & 0.00076 & 0.03721 & 93 & 178 & 882.47656 & 0.00103 & 0.12259 & 30 \\
24 & 873.33181 & 0.00078 & 0.04074 & 94 & 179 & 882.53052 & 0.00058 & 0.11770 & 32 \\
27 & 873.50795 & 0.00125 & 0.04033 & 24 & 195 & 883.47770 & 0.00127 & 0.12328 & 24 \\
28 & 873.56635 & 0.00101 & 0.03988 & 24 & 196 & 883.53627 & 0.00151 & 0.12300 & 23 \\
34 & 873.92186 & 0.00154 & 0.04229 & 13 & 208 & 884.24502 & 0.00016 & 0.12554 & 284 \\
39 & 874.22718 & 0.00091 & 0.05336 & 89 & 209 & 884.30611 & 0.00026 & 0.12779 & 321 \\
40 & 874.28617 & 0.00078 & 0.05350 & 93 & 212 & 884.48116 & 0.00066 & 0.12629 & 41 \\
43 & 874.46514 & 0.00217 & 0.05592 & 18 & 213 & 884.53689 & 0.00052 & 0.12317 & 47 \\
44 & 874.52324 & 0.00204 & 0.05517 & 24 & 226 & 885.30543 & 0.00028 & 0.12666 & 198 \\
45 & 874.57921 & 0.00146 & 0.05229 & 26 & 229 & 885.48206 & 0.00054 & 0.12674 & 39 \\
56 & 875.23046 & 0.00034 & 0.05619 & 95 & 230 & 885.53889 & 0.00094 & 0.12472 & 37 \\
57 & 875.29009 & 0.00050 & 0.05697 & 95 & 242 & 886.24842 & 0.00047 & 0.12805 & 203 \\
58 & 875.34864 & 0.00066 & 0.05668 & 93 & 246 & 886.48146 & 0.00112 & 0.12569 & 23 \\
60 & 875.46600 & 0.00110 & 0.05633 & 17 & 247 & 886.54157 & 0.00116 & 0.12695 & 28 \\
61 & 875.52677 & 0.00149 & 0.05825 & 20 & 263 & 887.48551 & 0.00187 & 0.12929 & 23 \\
62 & 875.58595 & 0.00065 & 0.05858 & 23 & 276 & 888.24735 & 0.00026 & 0.12608 & 236 \\
67 & 875.88340 & 0.00024 & 0.06178 & 59 & 293 & 889.25313 & 0.00032 & 0.13141 & 331 \\
69 & 876.00008 & 0.00055 & 0.06076 & 59 & 294 & 889.30584 & 0.00050 & 0.12527 & 220 \\
73 & 876.24152 & 0.00021 & 0.06680 & 792 & 310 & 890.25255 & 0.00039 & 0.13038 & 164 \\
77 & 876.47664 & 0.00109 & 0.06652 & 21 & 314 & 890.48615 & 0.00132 & 0.12858 & 23 \\
78 & 876.53781 & 0.00083 & 0.06884 & 20 & 315 & 890.54896 & 0.00281 & 0.13254 & 28 \\
79 & 876.59718 & 0.00172 & 0.06936 & 17 & 331 & 891.48948 & 0.00145 & 0.13146 & 22 \\
84 & 876.89396 & 0.00048 & 0.07189 & 59 & 338 & 891.89414 & 0.00085 & 0.12417 & 16 \\
85 & 876.95344 & 0.00043 & 0.07252 & 42 & 344 & 892.25418 & 0.00063 & 0.13111 & 173 \\
86 & 877.01135 & 0.00054 & 0.07158 & 59 & 348 & 892.48294 & 0.00084 & 0.12447 & 22 \\
91 & 877.31073 & 0.00048 & 0.07670 & 659 & 349 & 892.54779 & 0.00268 & 0.13047 & 24 \\
94 & 877.48801 & 0.00103 & 0.07744 & 20 & 365 & 893.48443 & 0.00138 & 0.12551 & 20 \\
95 & 877.54969 & 0.00153 & 0.08027 & 20 & 366 & 893.54403 & 0.00203 & 0.12626 & 18 \\
102 & 877.96326 & 0.00043 & 0.08189 & 59 & 372 & 893.89704 & 0.00047 & 0.12616 & 59 \\
111 & 878.49824 & 0.00107 & 0.08722 & 20 & 382 & 894.48629 & 0.00099 & 0.12692 & 20 \\
112 & 878.55744 & 0.00131 & 0.08757 & 22 & 383 & 894.54186 & 0.00181 & 0.12364 & 21 \\
128 & 879.50860 & 0.00102 & 0.09713 & 20 & 395 & 895.25137 & 0.00024 & 0.12695 & 216 \\
129 & 879.57040 & 0.00079 & 0.10008 & 22 & 399 & 895.48451 & 0.00175 & 0.12470 & 20 \\
134 & 879.86647 & 0.00030 & 0.10190 & 42 & 416 & 896.48726 & 0.00166 & 0.12698 & 20 \\
135 & 879.92464 & 0.00026 & 0.10122 & 59 & 433 & 897.48289 & 0.00242 & 0.12217 & 21 \\
136 & 879.98351 & 0.00051 & 0.10124 & 59 & 446 & 898.24512 & 0.00040 & 0.11935 & 340 \\
137 & 880.04119 & 0.00047 & 0.10007 & 59 & 447 & 898.30482 & 0.00211 & 0.12020 & 105 \\
140 & 880.21939 & 0.00026 & 0.10172 & 243 & 450 & 898.48273 & 0.00122 & 0.12156 & 21 \\
141 & 880.27973 & 0.00031 & 0.10321 & 145 & 458 & 898.94992 & 0.00060 & 0.11795 & 45 \\
142 & 880.33950 & 0.00064 & 0.10413 & 82 & 467 & 899.48192 & 0.00165 & 0.12030 & 19 \\
144 & 880.45810 & 0.00169 & 0.10503 & 13 & 480 & 900.24611 & 0.00032 & 0.11944 & 264 \\
145 & 880.51858 & 0.00044 & 0.10665 & 20 & 484 & 900.47802 & 0.00109 & 0.11595 & 20 \\
146 & 880.57851 & 0.00131 & 0.10774 & 21 & 497 & 901.24942 & 0.00127 & 0.12230 & 262 \\
151 & 880.87243 & 0.00026 & 0.10741 & 55 & 514 & 902.25082 & 0.00041 & 0.12325 & 234 \\
152 & 880.93282 & 0.00028 & 0.10895 & 59 & 518 & 902.48836 & 0.00096 & 0.12539 & 30 \\
153 & 880.99124 & 0.00040 & 0.10852 & 59 & 552 & 904.49857 & 0.00151 & 0.13470 & 22 \\
    \hline
    \multicolumn{10}{l}{\commenta BJD$-$2460000} \\
    \multicolumn{10}{l}{\commentb  $C= {\rm BJD~} 2460871.87867 + 0.05885 \times E$.} \\
     \multicolumn{10}{l}{\commentc Number of points used to determine the maximum.} \\	
     \end{tabular}
\end{table*}

%%%%%%%%%%%%%%%%%%%%%%%%%%%%%%%%%%%%%%%%%%%%%%%%%%

% Don't change these lines
\bsp	% typesetting comment
\label{lastpage}
\end{document}